\documentclass[nofootinbib,prl,twocolumn]{revtex4-1}
\usepackage{graphicx}
\usepackage{longtable}
\usepackage{amsmath}
\usepackage{amssymb}
\usepackage{subfigure}
\usepackage{hyperref}
\usepackage{stackengine}
\usepackage{color}

\newcommand{\mpl}{{M_{\rm {pl}}}}

\newcommand{\dd}{\, {\rm d}}

\newcommand{\lsim}{\;\mbox{\raisebox{-0.5ex}{$\stackrel{<}{\scriptstyle{\sim}}$}
}\;}

\begin{document}
\title{Early dark energy from massive neutrinos --- a natural resolution of the Hubble tension}
\author{Jeremy Sakstein}
\email[Email: ]{sakstein@physics.upenn.edu}
\affiliation{Center for Particle Cosmology, Department of Physics and Astronomy, University of Pennsylvania 209 S. 33rd St., Philadelphia, PA 19104, USA}
\author{Mark Trodden}
\email[Email: ]{trodden@physics.upenn.edu}
\affiliation{Center for Particle Cosmology, Department of Physics and Astronomy, University of Pennsylvania 209 S. 33rd St., Philadelphia, PA 19104, USA}

\begin{abstract}
The Hubble tension can be significantly eased if there is an early component of dark energy that becomes active around the time of matter-radiation equality. Early dark energy models suffer from a coincidence problem---the physics of matter-radiation equality and early dark energy are completely disconnected, so some degree of fine-tuning is needed in order for them to occur nearly simultaneously. In this paper we propose a natural explanation for this coincidence. If the early dark energy scalar couples to neutrinos then it receives a large injection of energy around the time that neutrinos become non-relativistic. This is precisely when their temperature is of order their mass, which, coincidentally, occurs around the time of matter-radiation equality. Neutrino decoupling therefore provides a natural trigger for early dark energy by displacing the field just before matter-radiation equality. We discuss various theoretical aspects of this proposal, potential observational signatures, and future directions for its study. 
\end{abstract}

\maketitle


The Hubble tension is one of the biggest mysteries confounding cosmologists today. The Hubble parameter derived by fitting the $\Lambda$CDM model to cosmic microwave background (CMB) data and its value measured locally using distance indicators and strong lensing time-delays are discrepant by more than $5\sigma$ \cite{Aylor:2018drw,Wong:2019kwg,Verde:2019ivm,Knox:2019rjx}, with the CMB favoring a lower value. The last five years have seen an immense effort to uncover any potential systematics in both the local \cite{Fitzpatrick:2000hh,Benedict:2006cp,Humphreys:2013eja,Efstathiou:2013via,Rigault:2014kaa,Becker:2015nya,Shanks:2018rka,Riess:2018kzi,Kenworthy:2019qwq} and CMB pipelines \cite{Spergel:2013rxa}, yet the tension persists after all reanalyses. Similarly, attempts to use alternate statistical measures of the tension have all reinforced this strong disagreement \cite{Bennett:2014tka,Cardona:2016ems,Zhang:2017aqn,Feeney:2017sgx}. Thus, the possibility that the discrepancy is due to new physics beyond the standard cosmological model has become an increasingly intriguing notion.

Theoretical explanations for the Hubble tension fall into two categories. {Local resolutions, where there is some systematic affecting the distance ladder}, include the possibility that we are located in an under dense void \cite{Lombriser:2019ahl} or that screened fifth-forces from modifications of gravity alter the calibration of the Cepheid period-luminosity relation \cite{Desmond:2019ygn,Sakstein:2019qgn,Desmond:2020wep}. In contrast, early universe resolutions, {which alter pre-recombination cosmology}, have focused on dark radiation \cite{Bernal:2016gxb,DiValentino:2016hlg,Vagnozzi:2019ezj}, strong neutrino self-interactions \cite{Kreisch:2019yzn}, and large primordial non-gaussianities \cite{Adhikari:2019fvb}. 

One particularly interesting resolution is early dark energy (EDE) \cite{Karwal:2016vyq,Mortsell:2018mfj,Poulin:2018cxd,Alexander:2019rsc,Niedermann:2019olb,Berghaus:2019cls}. In this scenario, an additional component of dark energy becomes active around the epoch of matter-radiation equality. During this period, the Hubble parameter decays at a slower rate than in $\Lambda$CDM. The sound horizon for acoustic waves in the photon-baryon fluid, given by
\begin{equation}
r_s=\int_{z_{\rm eq}}^\infty\frac{c_s}{H(z)}\dd z,
\end{equation}
is thus reduced compared with $\Lambda$CDM. The angular scale of the sound horizon at matter-radiation equality, $\theta_*$, is insensitive to this since it is determined solely by the location of the first CMB peak. This implies that the angular diameter distance to last scattering $D_A=r_s/\theta_*$, which depends inversely on the Hubble constant ($D_A\propto H_0^{-1}$), is also reduced, and therefore that $H_0$ is higher than in $\Lambda$CDM.

The success of the above scenario is crucially dependent on the EDE becoming active shortly before matter-radiation equality ($z_{\rm eq}\sim3000$) and rapidly becoming irrelevant thereafter. Had the EDE been active much before matter-radiation equality, its effect on the sound horizon would be negligible and the Hubble tension would persist. { Models where the EDE persists too long after matter-radiation equality are heavily disfavored by the data but the precise reason for this is poorly understood at present \cite{Knox:2019rjx}, and likely involves a complicated interplay of different competing cosmological processes. }

The simplest models of dark energy involve a scalar field slowly rolling down a potential \cite{Copeland:2006wr}. Initially, the field's mass is smaller than the Hubble parameter, so the scalar is overdamped by Hubble friction and remains at its initial position. At this point the field behaves as a sub-dominant cosmological constant. Once the Universe expands to the point where $H\sim m_\phi$, the driving force overcomes the friction and the field begins to slowly roll, during which time it acts as EDE. Finally, the field executes undamped oscillations where it behaves as a fluid with equation of state $w_\phi=(n-1)/(n+1)$ if the potential near the minimum scales like $V(\phi)\propto\phi^{2n}$.   

Theoretically, this scenario suffers from a coincidence problem. In order for the field to begin rolling around matter-radiation equality, its mass should be incredibly small, $m_\phi\sim 10^{-29}$ eV. Such a small mass is radiatively unstable unless all of its couplings to the standard model are fine-tuned to zero, or the field is the pseudo-Nambu-Goldstone boson of some spontaneously broken symmetry. The latter scenario can be realized if the field is an axion-like particle with a broken shift-symmetry, but in this case the leading-order potential is $V(\phi)=\Lambda^4[1-\cos(\phi/f)]$ so that $V(\phi)\propto \phi^2$ near the minimum and the scalar decays like pressureless matter after the EDE phase. This extra component of dark matter is strongly disfavored by supernovae measurements of the recent expansion history \cite{Poulin:2018dzj,Knox:2019rjx}. Potentials of the form $V(\phi)=\Lambda^4[1-\cos(\phi/f)]^n$ can alleviate the tension \cite{Poulin:2018cxd,Smith:2019ihp,Capparelli:2019rtn} { --- in fact, they fit the data better than power law models, the best fit being $n=3$ ---} but are themselves highly fine-tuned, since one is requiring higher-order instanton corrections to dominate over the leading-order terms. 

In this work, we propose an alternative mechanism to trigger the onset of EDE that is free of any fine-tuning. Rather than relying on balancing the scalar mass against the Hubble parameter, we exploit a natural coincidence in the energy scales at matter-radiation equality, namely that the temperature around this time is of order $1$ eV, tantalizingly close to the upper limit on the sum of the neutrino masses \cite{Aghanim:2018eyx,Aker:2019uuj}. {(More precisely, the neutrino temperature at $z=3000$ is $0.51$ eV.)} If the dark energy scalar is conformally coupled to neutrinos, then it experiences an energy injection around the time at which neutrinos become non-relativistic. This occurs precisely when their temperature is comparable to their mass; i.e. at matter-radiation equality. It is therefore possible for the field initially to lie at the minimum of its potential and then be displaced around matter-radiation equality by the neutrinos, thereby avoiding fine-tuning issues with its mass and initial condition. The scalar subsequently behaves as EDE as it begins to roll back towards the origin\footnote{The field need not lie strictly at its minimum in order for this mechanism to be effective. The field might initially be over-damped but then be kicked up the potential to a new position where its mass 
is larger than the Hubble parameter. 
The important point is that the initial mass need not be fine-tuned for this mechanism to operate. Here we focus on the case where the field is initially at its minimum for illustrative purposes.}. 

\begin{figure}
\centering
\includegraphics[width=0.35\textwidth]{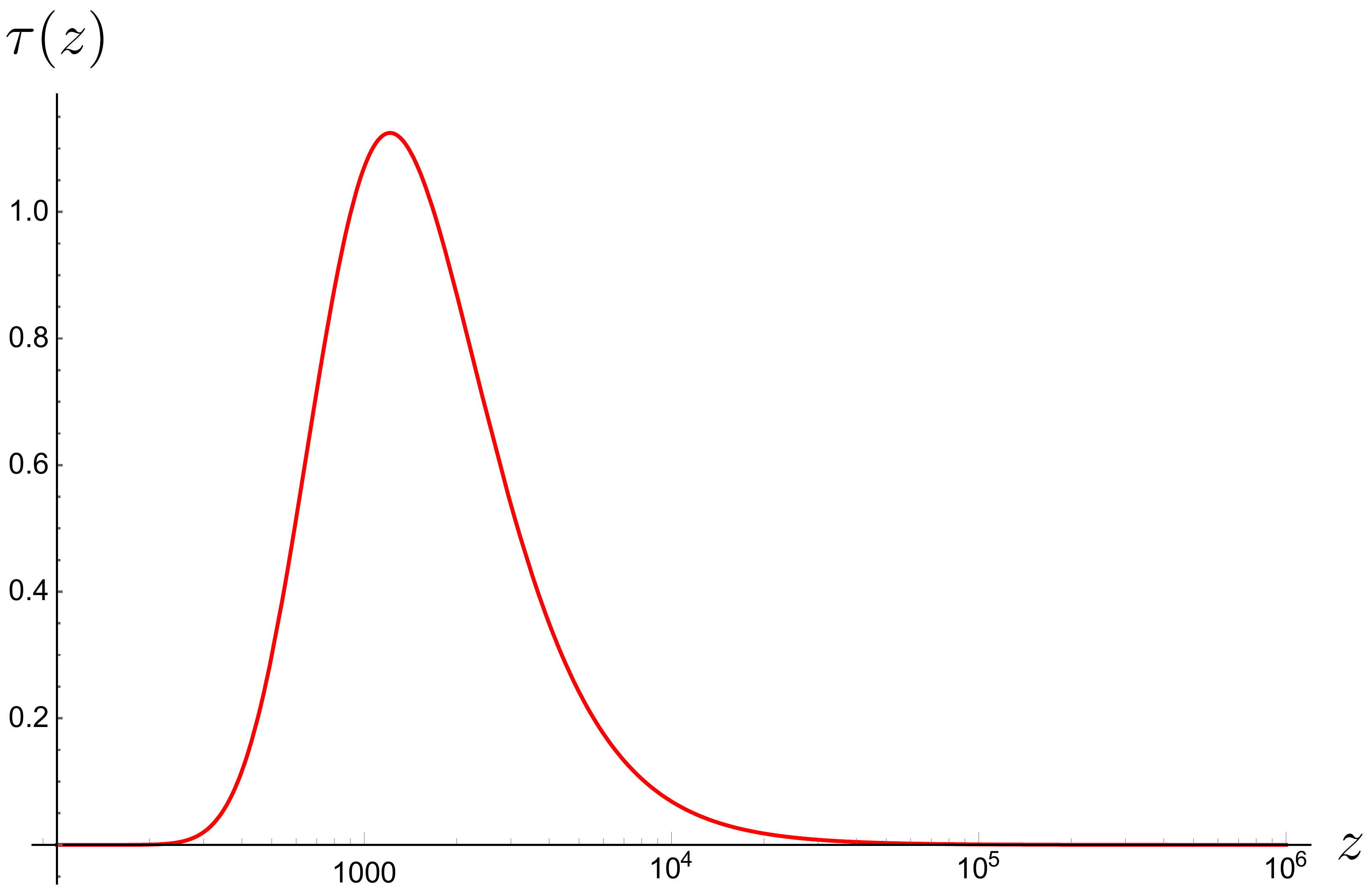}
\caption{The integral $\tau(m_\nu/T_\nu(z))$ as a function of $z$ with $m_\nu=0.5$eV.}
\label{fig:integral}
\end{figure}

To illustrate this mechanism, consider a simple model for EDE coupled to a single neutrino species (taken to be the heaviest of the three) via a conformal coupling to the metric $\tilde{g}_{\mu\nu}=e^{2\beta\frac{\phi}{\mpl}}g_{\mu\nu}$. (One could extend this theory to include extra couplings $\beta_i$ to different species, and to include mass-mixing, but here we focus on the simplest possible realization for illustrative purposes.)
The action is, 
\begin{align}\label{eq:act}
S&=\int\dd^4 x\sqrt{-g}\left[\frac{\mpl^2 R(g)}{2}-\frac12\nabla_\mu\phi\nabla^\mu\phi-V(\phi)\right]\nonumber\\&+S_\nu[\tilde g_{\mu\nu}],
\end{align}
where $S_\nu$ is the action for the neutrino sector but with all contractions made with $\tilde g_{\mu\nu}$ rather than $g_{\mu\nu}$. This can equivalently be written as
\begin{align}\label{eq:act2}
S&=\int\dd^4 x\sqrt{-g}\left[\frac{\mpl^2 R(g)}{2}-\frac12\nabla_\mu\phi\nabla^\mu\phi-V(\phi)\nonumber\right.\\&\left.+i\bar\nu\gamma^\mu\nabla_\mu\nu+m_\nu\left(1+\beta\frac{\phi}{\mpl}+\cdots\right)\bar\nu\nu\right],
\end{align}
where all contractions are now made with the metric $g_{\mu\nu}$. The equation of motion (EOM) in a Friedmann-Robertson-Walker background is 
\begin{equation}\label{eq:eom1}
\ddot{\phi}+3H\dot\phi+V'(\phi)=\frac{\beta}{\mpl}\Theta(\nu),
\end{equation}
where $\Theta(\nu)=g_{\mu\nu}\Theta(\nu)^{\mu\nu}$ is the trace of the neutrino energy-momentum tensor. Thus, the coupling to neutrinos can be viewed as contributing to an effective potential $V_{\rm eff}(\phi)=V(\phi)-\beta\Theta(\nu)\phi/\mpl$. Integrating over the Fermi-Dirac distribution, we obtain
\begin{align}
\Theta(\nu)&=-\rho_\nu+3P_\nu=-\frac{g_\nu T_\nu^4}{2\pi^2}\tau\left(\frac{m_\nu}{T_\nu}\right);\\&\tau(x)=x^2\int_x^\infty\frac{\left(u^2-x^2\right)^{\frac12}}{e^u+1}\dd u\label{eq:tauint},
\end{align}
where $g_\nu=4$ is the (massive) neutrino degeneracy and $T_\nu$ is the neutrino temperature, which is smaller than the photon temperature by a factor of $(4/11)^{\frac13}$ due to neutrino interactions having frozen out. The integral $\tau(x)$ is approximately zero when $x\gg1$, since the neutrino mass is negligible and it has equation of state $P\approx\rho/3$, and when $x\ll1$ due to Boltzmann suppression. When $x\approx 1$ ($T_\nu\approx m_\nu$) the integral is of order unity. This is depicted in figure \ref{fig:integral} where we plot $\tau$ as a function of redshift, showing that it peaks around matter-radiation equality. Let us suppose that the field is initially in its minimum\footnote{If the neutrino coupling is large enough, the minimum of the effective potential may differ significantly from the minimum of $V(\phi)$ so $\phi$ initially occupies this new minimum. The quantitative features presented in this work clearly persist in this case.}. Since the coupling acts as a forcing term in the EOM \eqref{eq:eom1}, the effect of the neutrino coupling is to kick the scalar out of its minimum and up its potential when $T_\nu\sim m_\nu$ \cite{Brax:2004qh,Coc:2006rt,Erickcek:2013oma,Erickcek:2013dea}. 

\begin{figure}[ht]
\centering
{\includegraphics[width=0.4\textwidth]{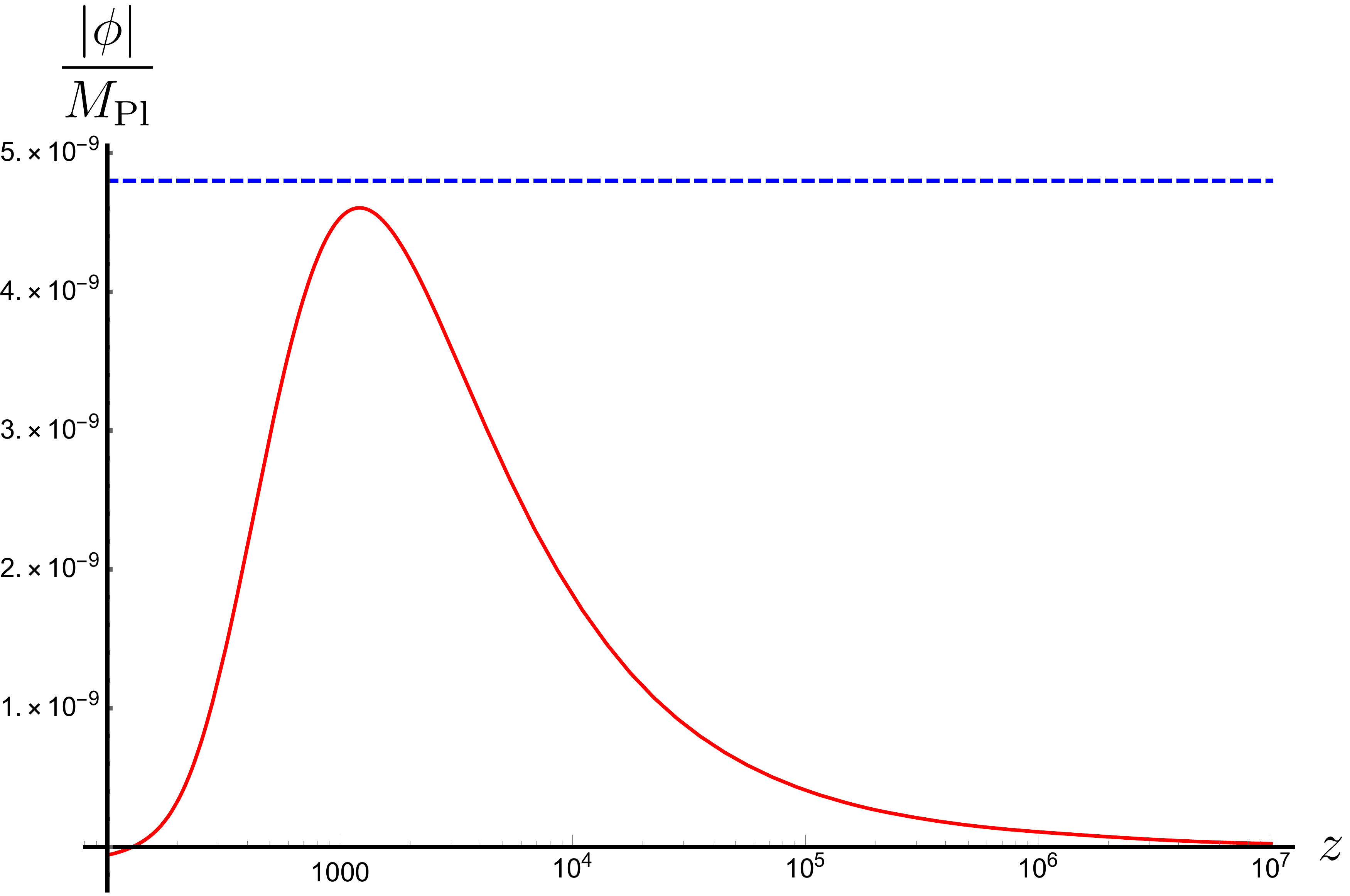}}
\caption{The field as a function of redshift (red, solid). The blue dashed line shows the analytic prediction in equation \eqref{eq:phiin}. We take $m_\nu=0.5$ eV, $\beta=4\times10^{-4}$, and $\lambda=10^{-75}$. }
\label{fig:results}
\end{figure}

We can estimate the magnitude of the kick as follows. First, we relate the neutrino temperature to the Hubble expansion via $3H^2\mpl^2=\pi^2/30 g_\star(T_\gamma)T_\gamma^4$ so that the EOM is
\begin{equation}\label{eq:EOM2}
\ddot{\phi}+3H\dot\phi+V'(\phi)=-\frac{45}{\pi^4}\left(\frac{4}{11}\right)^{\frac43}\frac{\beta g_\nu}{g_\star(T_\gamma)}H^2\mpl\tau\left(\frac{m_\nu}{T_\nu}\right).
\end{equation}
Let the temperature $T_\nu=m_\nu$ at time $t_k$. Since the integral is highly-peaked around this point, we can approximate the kick as a delta function so that $\tau(x)\approx 7\delta(t-t_k)/8H$, assuming that the energy is injected over a Hubble time. Neglecting the potential, we can then integrate equation \eqref{eq:EOM2} twice ($g_\star(1 \textrm{ eV})\approx 3.38$ and we assume the Universe is radiation-dominated) to find that $\phi$ is displaced from its initial location by an amount
\begin{equation}\label{eq:phiin}
\phi_k\approx -0.03\beta\mpl.
\end{equation}
This is the key result of our proposal. Equation \eqref{eq:phiin} is a natural initial condition for any EDE model where the scalar begins to roll shortly before matter-radiation equality. Furthermore, it is not necessary to fine-tune the mass to match the Hubble parameter around this time since the field is naturally displaced from the minimum due to its neutrino coupling. 

The novel features of our mechanism are insensitive to the precise form of the scalar potential but to explore further we will take $V(\phi)=\lambda \phi^4/4$. 
The action \eqref{eq:act} has an approximate scale-invariance broken by the neutrino mass term, so adding a scale-invariant potential is natural. Furthermore, it was shown in \cite{Agrawal:2019lmo} that this potential provides {a good} fit to the various data sets and can alleviate the Hubble tension by raising the derived value of $H_0$ to $72.3$ km/s/Mpc (at $2\sigma$). We have numerically solved the EOM \eqref{eq:EOM2} in conjunction with the Friedmann equations,
\begin{align}
3H^2\mpl^2&=\rho_m+\rho_\gamma+\frac{\dot\phi^2}{2}+\frac{\lambda}{4}\phi^4+\Lambda\mpl^2\\
\frac{\dot H}{H^2}&=-\frac{1}{2\mpl^2}\left(\sum_i (\rho_i+P_i) +{\dot\phi^2}{}\right),
\end{align}
where $i=\{m$,$\gamma \}$ and $\Lambda$ is the cosmological constant driving dark energy today. Representative results for $m_\nu=0.5$ eV (corresponding to the upper bound from Planck \cite{Aghanim:2018eyx} and assuming that the heaviest neutrino has a mass around this value),
$\beta=4\times10^{-4}$, and $\lambda=10^{-75}$ are shown in figure \ref{fig:results}. One can see the qualitative features discussed above are borne out. The field begins at its minimum in the early Universe, but when the temperature drops to values near the neutrino mass it is kicked up its potential to a value close to our analytic prediction in equation \eqref{eq:phiin}. Thereafter, Boltzmann suppression rapidly diminishes the driving term so the field falls back towards the minimum. The parameters were chosen to exemplify this scenario, and {one avenue for future research would be to determine the best-fitting potential and parameters using a full Markov Chain-Monte Carlo analysis, but this would require a rederivation of the neutrino Boltzmann hierarchy to include the EDE coupling. Such an analysis is beyond the scope of the present work, and will be performed separately.} 

Clearly, the qualitative features our mechanism will be similar for any choice of scalar potential, and so can be implemented into any of the EDE models that have been proposed. Another interesting possibility is to use the novel feature of energy injection into the scalar to construct alternative scenarios that cannot be achieved using quintessence-like models alone. To give one example, the kick from the neutrinos can be energetic enough to push the field over a local maximum in the potential. One can then envision a scenario where the field begins in a false vacuum, and, provided the lifetime of this minimum is long enough that tunneling does not occur, this field acts as EDE. When the temperature is of order the neutrino mass, the kick pushes the field over the local maximum, and, if the potential is steeper on the other side, the energy will rapidly dissipate. This scenario is certainly intriguing and it would be interesting to construct an explicit example in detail. This model-building exercise is postponed for future work. 


Our theory is an effective field theory (EFT) and requires a UV-completion. In particular, the $\mathrm{SU}(2)_L$ structure of the standard model is broken by our scalar-neutrino interaction, and the neutrino mass term. It is not difficult to construct UV-completions of these mass terms using, for example, additional Higgs fields, see-saw mechanisms, or supersymmetry. {It is also possible to do this in a technically natural manner \cite{DAmico:2016jbm,DAmico:2018hgc}.} Incorporating additional singlet scalars is trivial within this framework. It would certainly be interesting to embed our EFT within a more UV-complete model, and to connect it with the rest of the standard model, which would open up the possibility of testing our proposal at colliders e.g. using Higgs portal searches. 

We have taken the scalar to couple solely to neutrinos in order to study the simplest incarnation of our proposed mechanism, and it is possible to generalize the model to include couplings $\beta_i$ to other matter species, and possibly dark matter. Such couplings would not affect our mechanism. Their effects are three-fold. First, they induce extra kicks in the field at earlier times when particles more massive than neutrinos decouple. The effects of these kicks on the cosmology are largely irrelevant, but changes in the field value do cause the particle masses to vary by an amount $\delta m_i\sim \beta_i\Delta\phi$ so one needs to ensure that this shift is compatible with big bang nucleosynthesis, which constrains $\Delta m_i\lsim 10\%$ \cite{Brax:2004qh}. Second, the effective potential is now $V_{\rm eff}(\phi)=V(\phi)-\sum_i\beta_i\Theta(i)\phi$, which expands the number of possibilities for model building. Third, there are fifth-forces between particles proportional to $2\beta_i^2$. For dark matter, current constraints imply $\beta_{\rm DM}<0.1$ \cite{Kesden:2006zb,Kesden:2006vz} whereas the coupling to visible matter is constrained to be $\beta\lsim 10^{-3}$ \cite{Bertotti:2003rm,Sakstein:2017pqi}, but these can be circumvented if the model includes a screening mechanism \cite{Joyce:2014kja,Burrage:2017qrf}. These new features are highly model-dependent.

Coupling the scalar solely to neutrinos violates the equivalence principle (EP), and gives rise to an additional force between neutrinos of order $2\beta^2$ times the Newtonian force. This fifth-force/EP violation is difficult to test using laboratory methods. A more promising approach is to look for the effects of fifth-forces on cosmological observables such as the neutrino free-streaming length, which will alter the matter power spectrum at small scales \cite{Bird:2011rb}; and the clustering of neutrinos inside voids \cite{Massara:2015msa,Schuster:2019hyl}, which may be detectible with upcoming lensing surveys \cite{Banerjee:2019omr}.

Recently, strong neutrino self-interactions arising from a four-point interaction $\sim\widetilde G_F (\bar\nu\nu)(\bar\nu\nu)$ have been proposed as another potential resolution of the Hubble tension \cite{Kreisch:2019yzn,Escudero:2019gvw}. This has sparked thorough investigations of the CMB \cite{Cyr-Racine:2013jua}, astrophysical, and laboratory constraints \cite{Blinov:2019gcj} on such interactions, and, while there is some region of parameter space remaining, the vast majority is highly constrained. This effective four-point interaction arises by integrating out a heavy mediator in a more UV-complete theory. By necessity, our EDE scalar is massless at the minimum in order to avoid it contributing an extra dark matter component to the late-time Universe. For this reason, the constraints on the effective four-point operator do not apply. Instead, one must constrain the $2\rightarrow2$ scattering process {described by joining two cubic vertices that arise from the operator} $g\phi\bar\nu\nu$ in the action \eqref{eq:act2}. Here, $g=\beta(m_\nu/\mpl)\sim 10^{-27}\beta$. This is negligibly small unless $\beta$ is inconceivably large.  To quantify this, we can write $\beta=\mpl/\mathcal{M}$ in order to make the cut-off $\mathcal{M}$ for the EFT explicit \cite{Burrage:2017qrf}. A coupling $g\sim\mathcal{O}(1)$ implies a cut-off $\mathcal{M}\sim \mathcal{O}(\textrm{eV})$, which is incredibly low, and comparable to the scales at which we are using this theory to make predictions. 

Interestingly, our proposal is on the verge of being excluded, {and may even be so already}. Current constraints on the sum of the neutrino masses for the $\Lambda$CDM model yield $\sum m_\nu<0.54$ eV (solely Planck) and $\sum m_\nu<0.12$ eV (Planck + lensing + BAO) \cite{Vagnozzi:2017ovm,Aghanim:2018eyx}. { Other independent probes yield similar bounds \cite{Palanque-Delabrouille:2019iyz,Archidiacono:2020dvx}. Taken at face value, these bounds would exclude the simplest incarnation of our mechanism because the small mass splitting measured experimentally would imply that $\sum m_\nu\sim1.5$ eV. Such a conclusion is premature because these bounds apply strictly to $\Lambda$CDM, and it is has been demonstrated that adding non-standard neutrino interactions can increase the error bars significantly \cite{Kreisch:2019yzn}. To be fully consistent, one needs to perform an updated analysis accounting for the modified cosmology and neutrino sector on a model by model basis. We will develop specific models and derive the updated bounds in a forthcoming publication. If the simplest models are indeed disfavored then it may be possible to extend them to achieve consistency for the data. One such possibility is to use the fact that the neutrino mass varies as $m_\nu(\phi)=m_\nu(1+\beta\phi/\mpl)$ to find potentials that can both act as EDE and set the mass at later times to be compatible with observations. } 

To summarize, early dark energy is a promising resolution of the Hubble tension, but it suffers from a coincidence problem. One needs to fine-tune the model parameters so that it begins slightly before matter-radiation equality and ends rapidly thereafter. In this work, we have proposed a new mechanism to address this problem. If the EDE scalar is coupled to neutrinos then it receives an energy injection around the time that neutrinos become non-relativistic --- precisely around matter-radiation equality for an $\mathcal{O}(\textrm{eV})$ neutrino mass. Such a coupling  ameliorates the coincidence problem because the energy injection can displace the scalar up its potential from any initial condition, bringing about the onset of EDE without the need to tune the scalar's mass.

Our proposal lays the foundations for several follow-up research directions. Theoretically, there is much work to be done model building, and a full exploration of the types of potential that can now be accommodated using our mechanism is certainly warranted. It would also be interesting to construct explicit working examples of the field being pushed over maxima in the potential in order to dissipate the EDE  energy sufficiently rapidly. More realistic scenarios could be constructed by placing the mechanism into a more general framework, for example, studying the effects of coupling to multiple neutrino species and including the effects of neutrino mass-mixing. Similarly, incorporating the scalar into potential UV-completions of the neutrino mass sector may give rise to additional interactions with the standard model that could be used to test our proposal at colliders using, for example, Higgs portal couplings. Observationally, the scalar mediates a fifth-force between neutrinos, and a promising avenue of testing our proposal would be to calculate and constrain the effects of these on the matter power spectrum, and the clustering of neutrinos in voids. {These possibilities are currently under investigation, and we will report the results in forthcoming publications.}

{ Acknowledgements:} We thank Prateek Agrawal, Cora Dvorkin, JiJi Fan, Bhuvnesh Jain, Marc Kamionkowski, Tanvi Karwal, Qiuyue Liang, Sam McDermott, Evan McDonough, Marco Raveri, and Tristan Smith for insightful discussions. JS is supported by the Center for Particle Cosmology. The work of MT is supported in part by US Department of Energy (HEP) Award de-sc0013528, and by NASA ATP grant NNH17ZDA001N

\bibliography{ref}

\end{document}